\documentclass[12pt]{elsart}

\usepackage{cite,epsfig,amsfonts,amssymb}

\begin{document}
\runauthor{Carmona, Richert and Wagner}
\begin{frontmatter}

\title{Order parameter fluctuations and thermodynamic phase transitions in
finite spin systems and fragmenting nuclei}

\author[Zgza]{J. M. Carmona\thanksref{mail1}},
\author[Stbg1]{J. Richert\thanksref{mail2}},
\author[Stbg2]{P. Wagner\thanksref{mail3}}
\address[Zgza]{Departamento de F\'{\i}sica Te\'orica, Universidad de Zaragoza,
Pedro Cerbuna 12, 50009 Zaragoza, Spain}
\address[Stbg1]{Laboratoire de Physique Th\'eorique, Unit\'e de Recherche 
Mixte Universit\'e - CNRS UMR7085, Universit\'e Louis Pasteur,
 3, rue de l'Universit\'e 67084 Strasbourg Cedex, France}
\address[Stbg2]{Institut de Recherches Subatomiques, Unit\'e de Recherche 
Mixte Universit\'e - CNRS UMR7500, BP28, 67037 Strasbourg Cedex 2, France}
\thanks[mail1]{E-mail: jcarmona@posta.unizar.es} 
\thanks[mail2]{E-mail: richert@lpt1.u-strasbg.fr}
\thanks[mail3]{E-mail: pierre.wagner@ires.in2p3.fr}

\begin{abstract}
We show that in small and low density systems described by a lattice gas model 
with fixed number of particles the location of a thermodynamic phase transition
can be detected by means of the distribution of the fluctuations related to an 
order parameter which is chosen to be the size of the largest fragment. We 
show the correlation between the size of the system and the observed order of
the transition. We discuss the implications of this correlation on the analysis
of experimental fragmentation data.
\end{abstract}
 

\end{frontmatter}

 There exist by now several indications that nuclear matter can appear in
different phases which may be produced by means of energetic nuclear 
collisions. Signs for the possible existence of a phase transition have been
found experimentally through the construction 
of the caloric curve which relates
the temperature $T$ to the energy $E$ of the system, the extraction of the
specific heat and the analysis of the behaviour of fragment size distributions
\cite{Kreutz,Zheng,Pochodzalla,DAgostino,DAgostino2,Chbihi,Borderie,Elliott}.
Simple minded approaches like percolation and lattice gas models 
generate relevant quantities like fragment size distributions and 
thermodynamic observables whose features 
are characteristic for such transitions
\cite{Campi,Elattari,Campi2,Pan,Borg,Gulminelli,Chomaz,Carmona,Pleimling,Ma}.
Experiments are seemingly able to reproduce these features.
However, there remains a need for further experimental
confirmation and for clarification of some points as it will appear below.

 The aim of the present work is twofold. First we want to show that under the
assumption of thermodynamic equilibrium a judicious choice of observables 
related to fragment size distributions may be an efficient tool to detect a 
phase transition in small systems. Second we shall show that the order of the
transition which comes out of the present investigations can be different when
the system is small or large. Similar effects have been observed in other
systems \cite{Alonso}.

 Two facts will guide our investigations. First, Botet and Ploszajczak 
suggested very recently that the distribution of order parameter fluctuations 
can be helpful for the detection of a phase transition in a finite 
system, at least when the transition is continuous~\cite{Botet}. Such 
distributions show a different behaviour at and off a transition point and are 
scale invariant with respect to the size of the system at that 
point. An experimental test on Xe$+$Sn for different bombarding 
energies reveals indeed scaling properties of the measured 
events~\cite{Botet2}. 

 Second, there exists by now some hope and even hints that simple models like
lattice models and other related approaches work as generic frameworks which 
provide a realistic, even though possibly only qualitative description of 
nuclear fragmentation~\cite{Richert}. The lattice gas 
model (LGM)~\cite{Campi2} is the paradigm of this type of models. Its 
basic variables are the temperature $T$ and the density $\rho$ of particles.
Formulated in the grand canonical ensemble, it can be mapped onto the 
Ising model with a magnetic field $h$~\cite{Lee}, which, as it is well
known, presents a discontinuity in the equation of state involving the
magnetization, $M(T,h)$, for $T<T_c$ at $h=0$, where $T_c$ is the critical
temperature. This corresponds in the LGM to a first-order transition line
(the 'coexistence line') in the $(\rho,T)$ phase diagram, which separates
the homogeneous phases from the 'liquid' and 'gas' coexistence region.
On this line, there is a critical point at $\rho=0.5$, $T=T_c$.
The LGM in the canonical ensemble was considered in ref.~\cite{Gulminelli}.
In finite systems the previous discontinuity evolves into a backbending
in the chemical potential as a function of $\rho$ at subcritical temperatures,
which produces a negative branch in its derivative~\cite{Gulminelli}. 

 In this context we developed the so called IMFM (Ising Model with Fixed 
Magnetization)~\cite{Carmona}. The Hamiltonian reads 
\begin{displaymath} 
H_{\rm IMFM} = \sum_{i=1}^A \frac{p_i^2}{2m}+V_0 \sum_{<ij>} \sigma_i \sigma_j 
\end{displaymath}
where $A$ is the number of particles, $\{ \sigma_i = \pm 1 \}$ is related to 
the occupation of site $i$ through $s_i = (1 + \sigma_i)/2$, 
$\{ s_i = 0, 1 \}$, and the interaction acts between nearest neighbours $<ij>$.
The total number of particles $ \sum_i s_i = A$ is fixed, so that the
canonical partition function can be written as
\begin{displaymath} {\mathcal Z}(T) =
 \sum_{ \{ \sigma \} } {\rm e}^{- \frac{1}{T} H_{\rm IMFM} } 
\delta_{\sum_i s_i,A} \ .
\end{displaymath}
The determination of the order of the transition is however harder when
one considers the behaviour of observables as a function of $T$ while 
maintaining $A$ as a fixed parameter of the model. Both the specific
heat~\cite{Carmona} and the fragment distributions~\cite{Gulminelli,Carmona} 
present an apparent scaling for small systems, and some characteristic
features of a smooth transition~\cite{Carmona2}.

\begin{figure}[htb!]
\begin{center}
\epsfig{figure=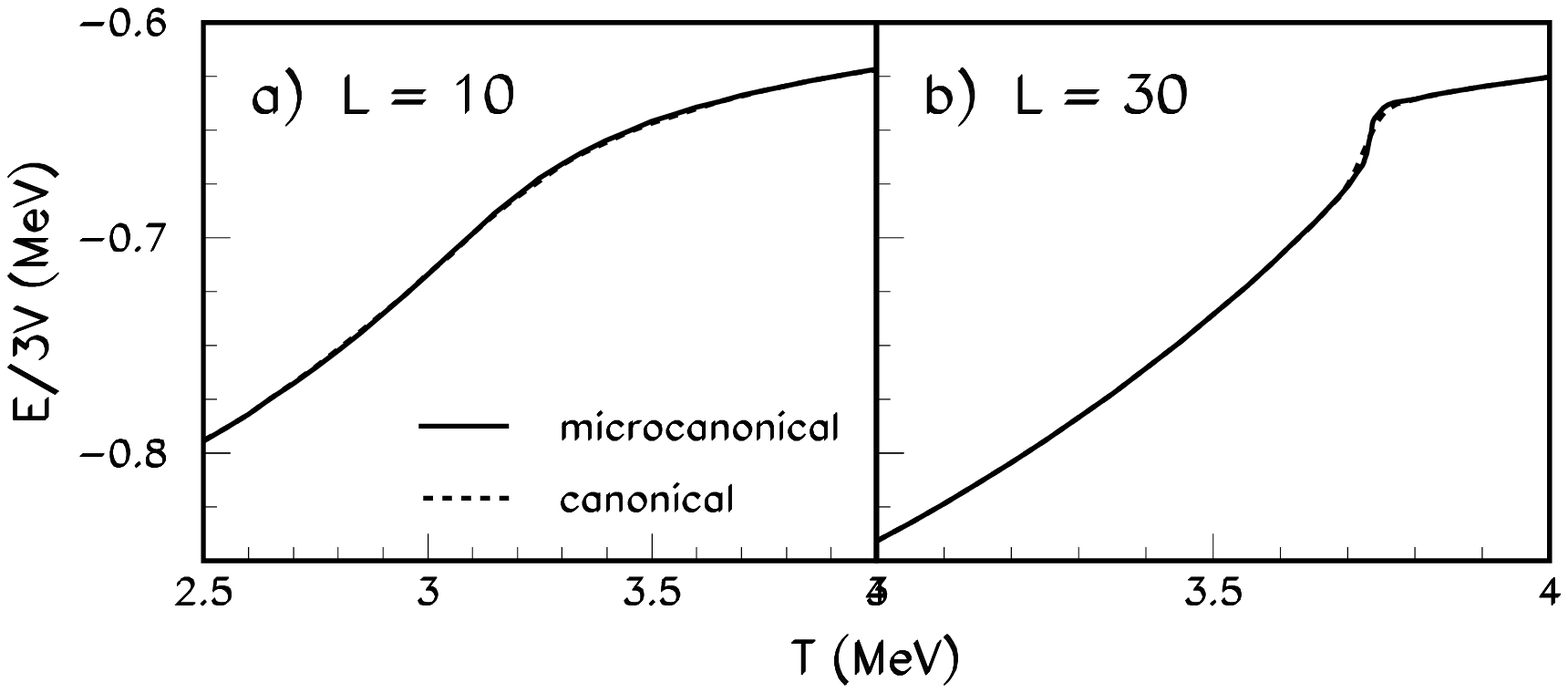,width=90mm}
\caption{Caloric curve $E$ vs. temperature $T$  for $3D$ systems with
a) $L=10$ and b) $L=30$ at density $\rho = 0.13$. The configuration energy
$E$ has been scaled by $3 V = 3 L^3$. The calculations are made in the 
framework of the IMFM model~\cite{Carmona}.}
\label{fig:1}
\epsfig{figure=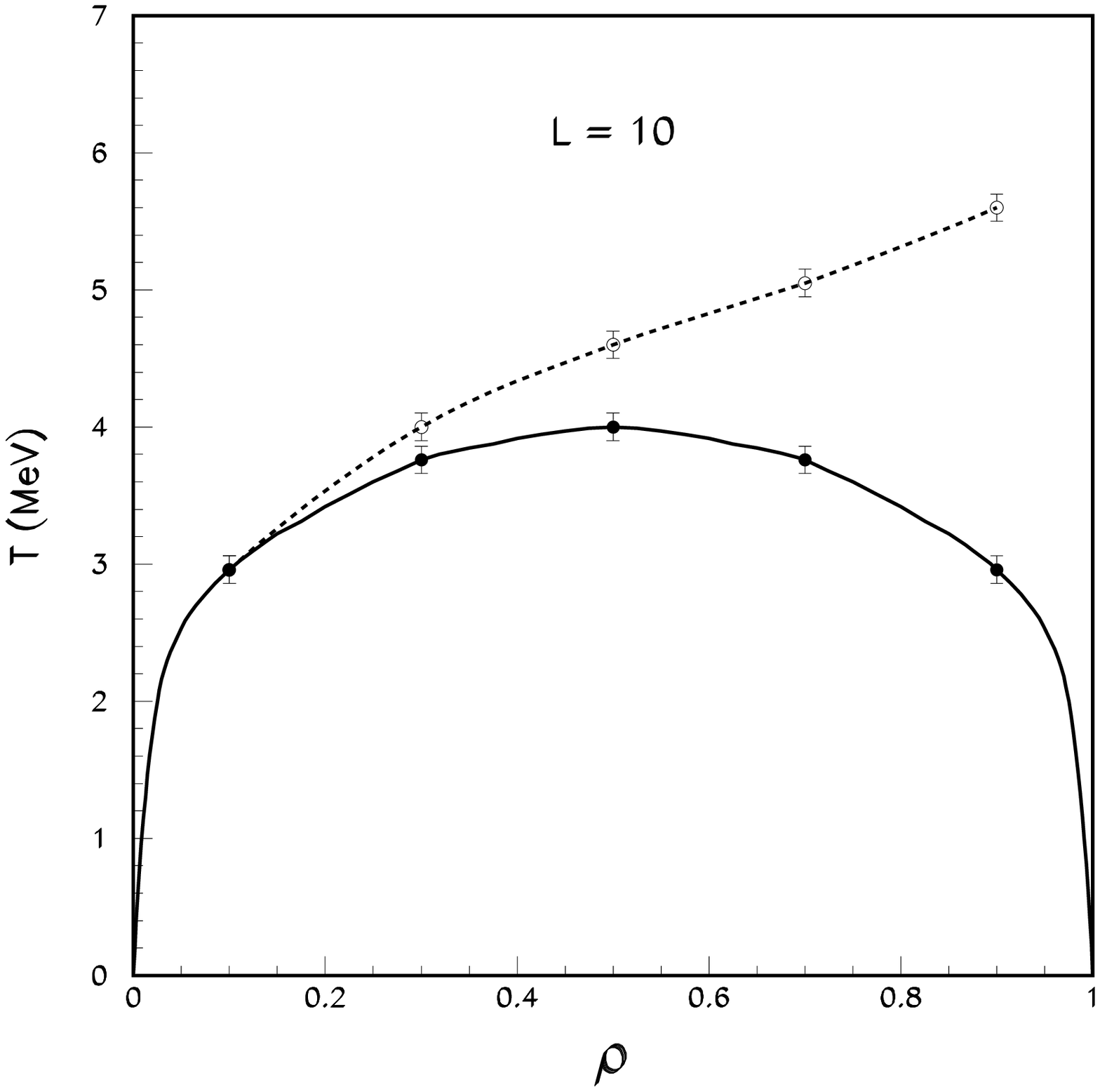,width=80mm}
\caption{Phase diagram in the $(\rho , T)$ plane. The solid line shows the
locus of $T_{tr}$, the dashed line the locus of $T_{f}$.}
\label{fig:2}
\end{center}
\end{figure}

Fig.~\ref{fig:1} shows typical caloric curves for $\rho = 0.13$ and 
different linear sizes $L$. Microcanonical 
and canonical calculations are indistinguishable up to $L = 30$. 
It is not possible to conclude about the order of the transition. This 
observation agrees with the microcanonical calculations of
Pleimling and H\"uller who showed that the first order nature of the 
transition appears only for $L \geq 60$~\cite{Pleimling2}.

 We have used the IMFM and systematically determined the values of the
transition temperature $T_{tr}$ corresponding to the maximum of the
specific heat and the temperature $T_f$ corresponding to the maximum of
the second moment of the fragment size distribution.
The specific heat does not diverge in the thermodynamic limit
(except for the critical point at $\rho=0.5$), but in finite systems 
it presents maxima which can indeed be observed
experimentally~\cite{DAgostino}. These maxima define a transition line
$(\rho,T_{tr})$ which however differs from the liquid-gas coexistence
line obtained by the usual Maxwell construction~\cite{Gulminelli}. 
In fact for finite systems it lies systematically below the coexistence 
line~\cite{Campi2,Gulminelli}, but both lines are expected to coincide in the 
thermodynamic limit~\cite{Pleimling2}. The use of $C_V$ is of interest 
in correlation with the fact that this quantity may be experimentally 
accessible~\cite{DAgostino,DAgostino2}.

 The locus of $T_{tr}$ and $T_f$ corresponds respectively to the 
full and the dashed line in the $(\rho , T)$ phase diagram shown in 
Fig.~\ref{fig:2}. For $\rho > 0.5$ $T_f$ is always larger 
than $T_{tr}$ but for $\rho < 0.5$ the temperatures $T_f$ and $T_{tr}$ 
come very close to each other, their distance decreases with 
decreasing $\rho$ and increasing $L$. Hence both maxima are correlated 
and different aspects of the same transition.

\begin{figure}[htb!]
\begin{center}
\epsfig{figure=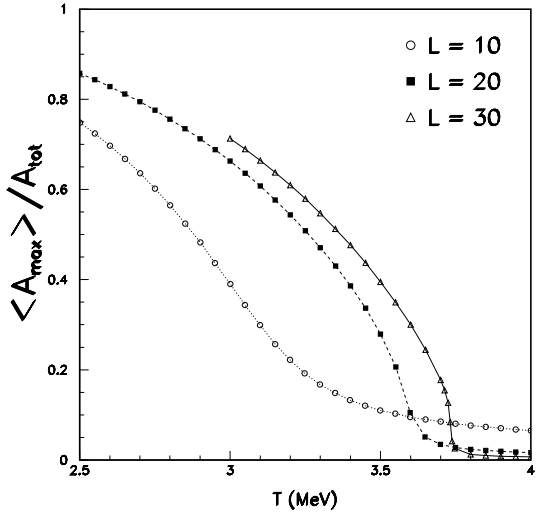,width=80mm}
\caption{Evolution of $<A_{max}>/A_{tot}$ ($A_{tot}$ = total number of 
particles) with the temperature for $\rho = 0.13$, $L = 10$ (open dots),
20 (squares), 30 (triangles). The calculations are performed in the 
framework of the microcanonical ensemble.}
\label{fig:3}
\end{center}
\end{figure}

Fig.~\ref{fig:3} shows the behaviour of the mean value of the size of the
largest cluster $<A_{max}>$ as a function of the temperature.
It appears that $<A_{max}>$ like the second moment varies more or less abruptly
in an interval of temperatures which lies close to $T_{tr}$.
If the IMFM provides a realistic picture of the underlying physics of nuclear
fragmentation it comes out that for low densities $\rho$ the observable
$A_{max}$ can be used as an order parameter which signals the presence of a 
thermodynamic phase separation point. We use $A_{max}$ in order 
to implement scaling tests. Following the arguments of ref.~\cite{Botet} we 
consider the function
\begin{displaymath} 
\Phi (z) \equiv  
\Phi \bigg( \frac{A_{max} - A_{max}^*}{<A_{max}>^\Delta} \bigg)
\equiv <A_{max}>^\Delta P(A_{max})
\end{displaymath}
where $A_{max}^*$ is the most probable value of $A_{max}$, 
$\Delta$ a real positive number and $P(A_{max})$ the normalised
probability distribution function of $A_{max}$. At a continuous transition 
point and $\Delta = 1$ the distribution $\Phi$ which shows the properties of 
the fluctuations of the order parameter $A_{max}$ is a scale invariant 
quantity.

\begin{figure}[htb!]
\begin{center}
\epsfig{figure=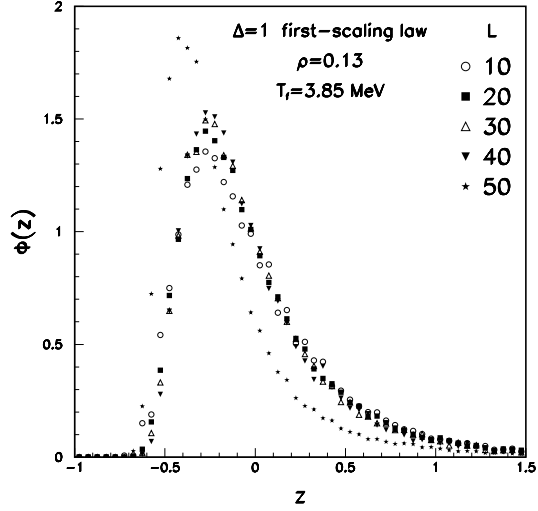,width=80mm}
\caption{The scaling behaviour of the fluctuations of $A_{max}$ for systems
of different linear sizes $L$ at $T_f = 3.85$ MeV and density $\rho = 0.13$. 
Here $\Delta = 1$. The calculations are performed in the canonical ensemble.}
\label{fig:4}
\end{center}
\end{figure}

\begin{figure}[htb!]
\begin{center}
\epsfig{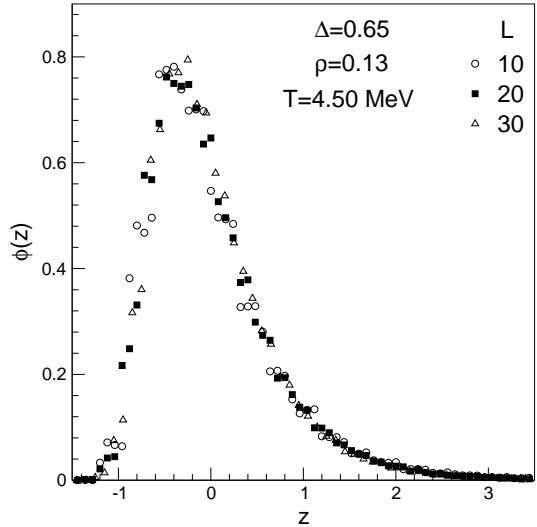}
\caption{Scaling behaviour of the systems of size $L = 10, 20, 30$,
$\rho = 0.13$, $T = 4.5$ MeV. Here $\Delta = 0.65$. The calculations
are performed in the canonical ensemble.}
\label{fig:5}
\end{center}
\end{figure}

 This is indeed the case in our model for systems of linear size $L \le 40$ 
as it can be seen in Fig.~\ref{fig:4}. The functions $\Phi$ show the 
characteristic scale invariance of the fluctuations for systems with 
different linear sizes $L$ at a temperature which corresponds
to the value of $T_{tr}$ for the infinite system.
Above $T_f$ scale invariance can be observed for $\Delta \neq 1$, even though
the overlap between the scaling functions for different values of $L$ is not
perfect as it can be seen in Fig.~\ref{fig:5}.

 In practice, events are collected with a certain experimental width in
energies or temperatures. One can then ask whether the scaling signature
would survive in a real experiment. In order to evaluate this effect, we
simulated systems with $L=$ 10, 20, 30, and 40 in a range of temperatures
$T=3.85\pm 0.20$ MeV at $\rho=0.13$ and used these data to produce $\Phi(z)$.
We observed that the $L=10$ and $L=20$ data still lie on the same curve,
while this is no longer true for $L\geq 30$. The result is not surprising, 
the smallest systems have a wider transition region and are consequently less
sensitive to shifts of the temperature. However, this limitation is not crucial
in practice. A system with $L=10$ at $\rho=0.3$ corresponds to 300 particles;
$L=20$ at $\rho=0.13$ corresponds to 1040 particles. In fact the scaling
signature has already been observed in collision experiments of Xe$+$Sn
(around 245 particles)~\cite{Botet2}. In an experiment with much larger
number particles and a large temperature dispersion, the signal would
effectively be lost.

 Fits of the tails of the fluctuation distributions on the transition line
are in agreement with the parametrization  
\begin{displaymath} \Phi(z) = a \exp(-bz^{\tilde{\nu}}) .\end {displaymath}
At $T_f = 4.5$ MeV one gets $\tilde{\nu} = 1.78 \pm 0.20$ for $\rho=0.50$ 
and $\tilde{\nu} = 1.15 \pm 0.15$ for $\rho = 0.30$ which may be 
compared with $\tilde{\nu} = 1.6 \pm 0.4$ obtained through the experimental 
analysis of ref.~\cite{Botet2}.  

\vspace{0.8cm}

 The present analysis shows that the generation of largest fragment fluctuation
distributions may be an efficient tool to detect a thermodynamic phase
transition in small finite systems of low density. Fragmentation events which
show scaling properties with respect to the size of the system correspond to 
events which lie at or in the near neighbourhood of such a transition. They
could be used as a tool in an experimental analysis and allow to detect a 
thermodynamic phase transition in the $(\rho, T)$ plane by selecting events 
with scaling properties and correlating them with thermodynamic observables 
like temperature and density. 

 There remains however an open point. The tools proposed in ref.~\cite{Botet} 
work for continuous transitions. On the other
hand the thermodynamic transition is a first order transition as indicated by
simulations of large systems. This can be observed in the framework of the 
microcanonical ensemble~\cite{Pleimling2}. It may sound contradictory that the 
transition related to the behaviour of the fragment size distribution and the 
thermodynamic transition are of different nature. In order to investigate this
point we extended our simulations to systems of size $L=50$ and 60, in the
framework of the canonical and microcanonical ensemble. The first order nature
of the transition reveals itself in both ensembles. In the case of the 
microcanonical approach our results confirm those of ref.~\cite{Pleimling2}.
In the canonical case, the first order character can clearly be seen. Indeed
in the neighbourhood of the thermodynamic transition, the energy distribution
shows the characteristic double hump which distinguishes the coexistence
of two phases. Working out the distribution of fluctuations 
$\Phi(z)$ one observes that this observable does no longer scale with the 
size of the system for $L \ge 50$ as it can be seen in Fig.~\ref{fig:4}.
This also happens with the $\Phi(z)$ derived from microcanonical calculations.

 The IMFM offers a possible explanation for the fact that one observes 
on the one hand features which are consistent with percolation 
concepts~\cite{Zheng} and, on the other hand, 
a first-order thermodynamic transition. 
The separation line defined by $T_f$ lies above the thermodynamic transition
for $\rho > 0.5$~\cite{Campi2}. But for $\rho<0.5$ densities, the fragment
distribution is intimately related to the thermodynamic behaviour
as it is in Fisher's phenomenological droplet model~\cite{Elliott}. Hence,
if the freeze-out happens at low densities, fragment formation is controlled
by the thermodynamic transition. The scaling observed in 
experiments could be due to the apparent, transient, continuous behaviour 
which this transition presents for small systems ($L \lesssim 50, 60$) in 
the framework of the IMFM. One should however notice that
negative values of $C_V$ which should corroborate the observation of a first
order transition have been seen experimentally~\cite{DAgostino2,DAgostino3} 
in as small systems as nuclei, in agreement with a specific microcanonical
treatment of the liquid-gas phase transition~\cite{Chomaz}.
  
 These observations lead to several conclusions. First, different observables 
may lead to different transition orders if one deals with small systems, 
and the order may change with the size of the system. Second, in the present 
case, the canonical and microcanonical treatments lead to the same 
answer for both small and large systems if one considers the pertinent 
quantity, here the energy distribution in the vicinity of the transition 
point. One should notice that this is not always the case. The
nonequivalence between microcanonical and canonical ensembles has been
proven in systems with negative specific heat regimes, corresponding
to canonically unstable states~\cite{canonical}. Analogously,
the backbending in the chemical potential implies nonequivalence
between the canonical and grand canonical descriptions of the LGM.
In fact this backbending reflects the first order character of the 
transition~\cite{Gulminelli}. We were interested however in the behaviour of
two other quantities, the fragment 
distributions and the specific heat, which are both 
experimentally accessible~\cite{DAgostino,DAgostino2}.
Third, the present behaviour of $\Phi(z)$ seems to indicate that 
scaling works in the case of a continuous transition as predicted in 
ref.~\cite{Botet} but not if the transition is first order. Fourth, 
one observes a consistency between the behaviour of $\Phi(z)$ and the
correct thermodynamic limit. The apparent scaling works only when the
system is small. When the system is large the first order behaviour
reveals itself in this observable and scaling fails.

 In summary, we have shown the correlation in the IMFM between
observables related to fragment size distributions and a thermodynamic 
transition. For small systems the distribution function of the 
largest fragment fluctuations shows the scaling features which may 
characterise a continuous phase transition. 
For systems much larger than nuclei, the first 
order character of the transition appears and the scaling properties of the 
distribution of the order parameter $A_{max}$ disappear.

\begin{ack}
 The authors acknowledge an interesting discussion with M.~Ploszajczak and 
R.~Botet. The work of JMC was partially supported by EU TMR program  
ERBFMRX-CT97-0122.
\end{ack}

\end{document}